# RADIO RESOURCE MANAGEMENT FOR EICIC, COMP, RELAYING AND BACK-HAULS TECHNIQUES IN LTE-ADVANCED NETWORKS: CONCEPTS AND A LITERATURE SURVEY

Najem N. Sirhan and Manel Martinez-Ramon

Electrical and Computer Engineering Department,
University of New Mexico, Albuquerque, New Mexico, USA

## ABSTRACT

*Frequency reuse in cells is one of the strategies that LTE (Long Term Evolution) uses to maximize the spectrum efficiency. However, it leads to an interference among the cells, especially at the cell edges where the probability for a cell-edge user to be scheduled on a resource block that is being transmitted by the neighbouring cell is high; consequently, the interference is high. In-order to mitigate Inter-Cell Interference (ICI), Inter-Cell Interference Coordination (ICIC) was proposed by the Third Generation Partnership Project (3GPP) standards for the LTE network, and later on, the enhanced Inter-Cell Interference Coordination (eICIC) was proposed for the LTE-Advanced network. ICIC reduces cell-edge interference on traffic channels from neighbouring cells by the use of three interference reduction schemes that works in the power and frequency domain, and they are based on lowering the power of some channels to limit their reception to the users that are close to the base station, and by reducing the chance of frequency overlap. eICIC was proposed to handle ICI in Heterogeneous Network (HetNet) deployments. It reduces the interference on both the traffic and control channels. It uses power, frequency and also time domain to mitigate intra-frequency interference in HetNets. Another technique that was proposed by the 3GPP for the LTE-Advanced network that can reduce the ICI and improve the cell average and cell-edge user throughput is the Coordinated Multi-Point (CoMP) transmission/reception technique. CoMP can increase the cell average and cell edge user throughput in both the uplink and downlink transmission by joint scheduling and data processing in multiple cells/eNBs. Another technique that was proposed by the 3GPP for the LTE-Advanced network that can improve the LTE network coverage in difficult conditions is to deploy Relay Nodes (RNs). In this paper, we survey Radio Resource Management (RRM) for some of the techniques that are used with LTE-A. The included techniques in this paper are; the ICIC, eICIC, CoMP, Relaying and Back-hauls. We start by explaining the concepts of these techniques. Then, by summarizing the radio resource management approaches that were proposed in the literature for these techniques. And finally, we provide some concluding remarks in the last section.*

## 1. INTRODUCTION

Long Term Evolution (LTE) and LTE-Advanced (LTE-A) are defined by the Third Generation Partnership Project (3GPP) standards as Releases 8/9 and Releases 10/11 respectively. The number of LTE (Long-Term Evolution) users and their applications has increased significantly in the last decade, which increased the demand on the mobile network. LTE-Advanced (LTE-A) comes with many techniques that can support and fulfil this increasing demand. These techniques are; the enhanced Inter-cell Interference Coordination (eICIC), Coordinated Multi-point operation (CoMP), Relaying and Back-hauls [16][17][18].

Radio Resource Management (RRM) refers to the whole functionality of managing the use of radio channels, co-channel interference, and other radio transmission characteristics. This management functionality is accomplished by the use of intelligent strategies and algorithms that include in their calculation multiple parameters in order to optimize the use of the available radio channels. These parameters include but are not limited to, the number of users, transmission rate per user, transmission power, Quality of Service (QoS) parameters, number of available channels, channels' conditions, and modulation and coding scheme [19] [21].

LTE uses Orthogonal Frequency Division Multiplexing (OFDM) as the basic signal format, that is a method which can successfully eliminates the effect of intra-cell interference by enabling the users in each cell to transmit orthogonally [20]. However, the effect of inter-cell interference will be present in the case of two cell-edge users are located in two adjacent cells communicating at the same frequency or in the case of these two users are causing interference to each other because of the high-power level at which they are transmitting. Inter-Cell Interference (ICI) will reduce the system performance in terms of signal to interference and noise ratio (SINR) values, system capacity, users' potential data rates [22]. In-order to deal with cell-edge interference, 3GPP introduced Inter-Cell Interference Coordination (ICIC) in LTE Release 8/9 (Rel-8/9). ICIC is considered a great technique that works well in a homogeneous network deployment, but it fails in a HetNets deployment. This is because LTE Rel-8/9 supports only the non-overlapping transmissions in the frequency domain. In LTE-Advanced, the enhanced Inter-Cell Interference Coordination (eICIC) was introduced in-order to manage the interference issues in HetNets. eICIC reduces the interference on both the traffic and control channels. It uses power, frequency and also time domain to mitigate intra-frequency interference in HetNets [21].

Another technique that was proposed by the 3GPP in LTE-Advanced system is the Coordinated Multi-Point (CoMP) transmission/reception technique, which increases the cell average and cell edge user throughput in both the uplink and downlink transmission. In CoMP multiple base stations (eNodeBs) cooperate to provide services to a single user or multiple users by what is called the joint scheduling scheme, so the interferences that are coming from adjacent base station will be changed to a useful signal after signal combination at the User Equipment (UE) side [6]. Another technique that was proposed by the 3GPP in LTE-Advanced system that can improve the LTE network coverage in difficult conditions, such as to improve urban or indoor throughput, or to extend coverage in rural areas, or to add dead zone coverage is to deploy Relay Nodes (RNs) [3].

The rest of the paper is structured as follows. The concept of the Inter-cell Interference Coordination (ICIC) is explained in section 2, the concept of enhanced ICIC (eICIC) is explained in section 3, the concept of the Coordinated Multi-point operation (CoMP) is explained in section 4, and the concept of relaying and back-hauls is explained in section 5. A summary of some the proposed radio resource management approaches for ICIC in section 6. A summary of some the proposed radio resource management approaches for CoMP in section 7. A summary of some of the proposed radio resource management approaches for relays in section 8. And finally, we provide some concluding remarks in the last section.

## 2. INTER-CELL INTERFERENCE COORDINATION (ICIC)

One of the main features of LTE that maximizes the spectrum efficiency is frequency reuse. It allows cells to use the same frequency channels, which will lead to an interference among the cells, especially at the cell edges. At the cell edges, the probability for a cell-edge user to be scheduled on a resource block which is being transmitted by the neighbouring cell is high; consequently, the interference is high. In-order to deal with cell-edge interference, 3GPP Release 8/9 (Rel-8/9) introduced Inter-Cell Interference Coordination (ICIC). ICIC reduces cell-edge interference on traffic channels e.g. Physical Downlink Shared Channel (PDSCH) from neighbouring cells, and this is done by the use of three interference reduction schemes that works in the power and frequency domain. These schemes are based on reducing the chance of frequency overlap. The first scheme, is by instructing the two neighbouring eNodeBs to use completely different sets of resource blocks throughout the cell at a given time. This scheme will significantly reduce interference, but it comes with a cost of not fully utilizing the whole set of resource blocks. In the second scheme, the eNodeBs distinguishes between centrally and edge located users, in this way eNodeBs can utilize the whole set of resource blocks for centrally located users, but for the edge users it uses completely different set of resource blocks. The third scheme is an enhanced version of the second one. The resource blocks are used in the same exact way, but all neighbouring eNodeBs uses different power schemes for their edge located users [21].

The flexibility nature of the PDSCH in the in LTE frame allows the ICIC schemes to directly work on scheduling them. However, in the case of the Physical Downlink Control Channel (PDCCH), ICIC schemes can't work directly on them, because they have a very different channel structure and they are much less flexible than the PDSCH. Hence, ICIC Interference reduction schemes can be applied only on traffic channels, and they can't be applied on control channels [21].

## 3. ENHANCED ICIC (EICIC)

Since Rel-8/9 supports only non-overlapping transmissions in the frequency domain, ICIC works well in a homogeneous networks environment, but fails in a HetNets environment. In 3GPP Release 10/11 (Rel-10/11), the enhanced Inter-Cell Interference Coordination (eICIC) was introduced in-order to manage the interference issues in HetNets. eICIC reduces the interference on both the traffic and control channels. It uses power, frequency and also time domain to mitigate intra-frequency interference in HetNets. eICIC has two main features: Cell Rang Expansion (CRE), which was introduced in 3GPP Release 8, and then enhanced in 3GPP Release 10/11. And the other feature is the Almost Blank Sub-frame (ABS), which was introduced in 3GPP Release 10 [21].

CRE allows the coverage of a pico-cell or femto-cell to expand in-order to include more users which exists at its edges. In 3GPP Release 8, the process of selecting the cell in which its range will be expanded is based on the received signal strength, but this approach is limited to the order of up to 9 dB gain. In 3GPP Release 10, the process of selecting the cell is based on minimum path loss and the interference levels [21].

ABS is a time-domain-based eICIC, it is a Time Domain Multiplexing (TDM) technique, it allows both the macro-cell and the small-cell "whether it is a pico-cell or femto-cell" to transmit over the same radio resources, however, the transmission is done at a different time slots. This is done by muting certain sub-frames of one layer of cells in-order to lower the interference in the other layer as shown in Figure 1 [21].

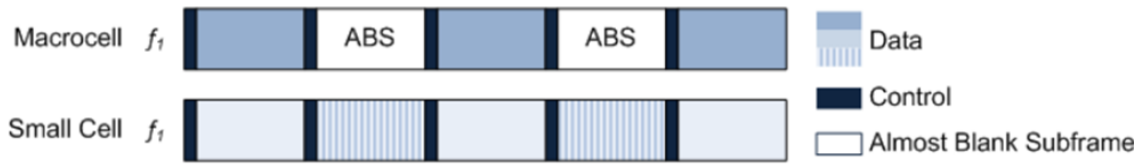

Figure 1. The concept of Almost Blank Sub-frame (ABS) [21]

## 4. COORDINATED MULTI-POINT OPERATION (COMP)

CoMP could be described as the art of interference management. If the transmitted signals from different cells are coordinated, the user's performance will be improved especially at cell-edges. An example of CoMP is shown in Figure 2. Two main categories are there for CoMP; the Inter-Site CoMP, and the Intra-Site CoMP. In the case of the Intra-site CoMP, it enables the coordination between sectors of the same Base Stations "eNodeBs". The coordination is possible through the use of Multiple Antenna Units (MAUs) that allow the coordination between the sectors. On the other hand, Inter-site CoMP enables the coordination between different BSs [12].

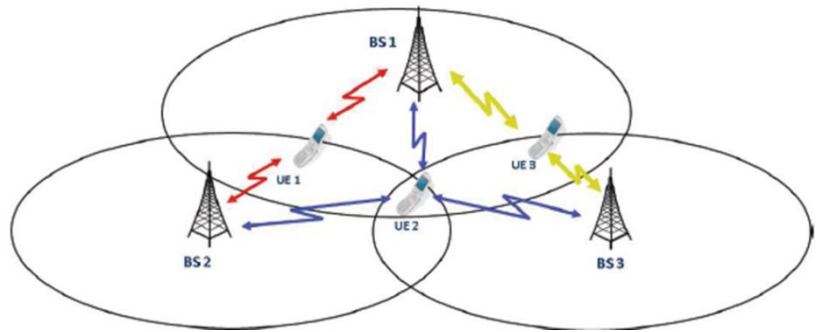

Figure 2. An example of three base stations (eNodeBs) coordinating among them to provide a better service for UE2 "cell-edge user" [12]

The two main approaches that are being used for coordination, are the centralized coordination as shown in Figure 3, and the distributed coordination as shown in Figure 4.

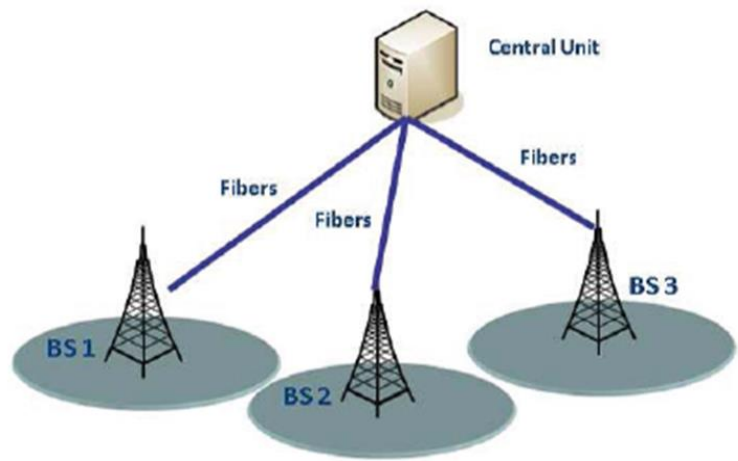

Figure 3. An example of a centralised CoMP, in which a server central unit is connected to three base stations through optical fibre links [12]

In the Centralized Coordination, the feedback and Channel State Information (CSI) data are available and processed at a central unit where it is responsible for handling Inter Cell Interference (ICI) and radio resource scheduling, processed data are sent to the coordinated cells over a star network. The main issue that arises in the use of this architecture is the high back-haul overhead and the stringent latency requirements [12].

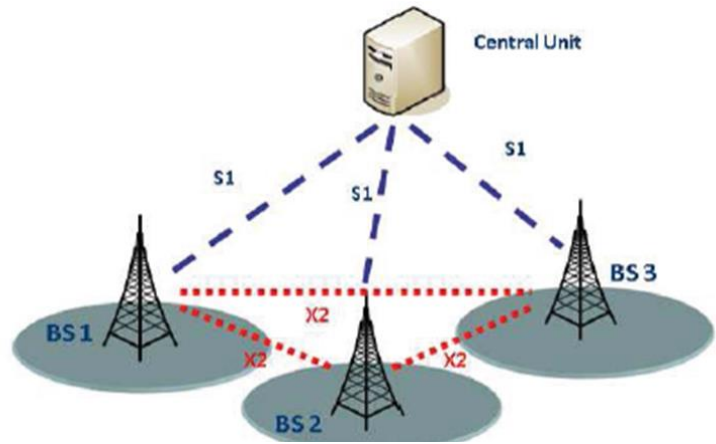

Figure 4. An example of distributed CoMP, in which the three base stations are connected to each other through an X2 interface in a mesh topology, in addition to being connected to a server central unit through an S1 interface [12]

In the Distributed Coordination, the feedback and Channel Signal Indicator (CSI) data are being exchanged between the coordinated cells over a fully meshed signalling Network using X2 interface, one of the coordinated cells can operate as a master cell while the other nodes as slaves, the master cell perform a central scheduling that manages resource allocation and re-transmission in a coordinated manner [12].

In downlink, there are two schemes for CoMP; the first one is the Coordinated Scheduling and/or Beam-Forming (CS/CB), and the second one is the Joint Processing (JP). Both of these downlink schemes are good solutions to mitigate ICI in the downlink of Multiple Input Multiple Output (MIMO) Orthogonal Frequency Division Multiplexing (OFDM) systems [12].

Coordinated Scheduling/Beam-forming (CS/CB) is characterized as a combination of multiple joint base stations coordination and Dynamic Inter Cell Interference Coordination (D-ICIC) schemes. CS/CB uses the MIMO antenna capabilities through beam-forming in a coordinated manner. While using beam-forming scheme, beams of different cells might collide as shown in Figure 5. Therefore, neighboring cells have to guess the interference that will be experienced. In order to face this problem the Coordinated Beam-Switching (CBS-CoMP) and the Coordinated Scheduling (CS-CoMP) were proposed. In the case of the CBS-CoMP, each cell determines a sequence of beams over which it continuously cycles. Coordination could be distributed between cell sites, or it could be centralized through a master scheduler. In the case of the CS-CoMP, ICI is being mitigated by enabling the collaboration of multiple base stations. The most famous CS-CoMP schemes enable the Coordination of multi-cell Pre-coding Matrix Index (PMI) between cooperating base stations, which allows the option of not using a centralized scheduler [12].

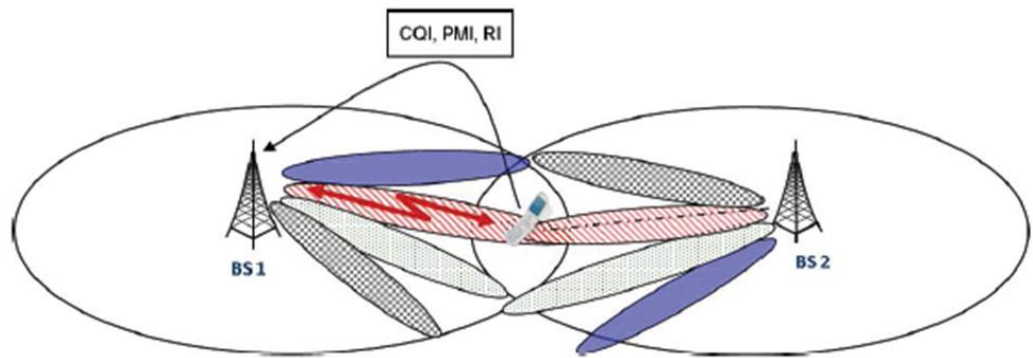

Figure 5. An example of UE feedback and beam collision in an uncoordinated beam-forming scenario [12]

In Joint Processing (JP), a CoMP set consists of a number of cell-sites that coordinate to optimize the cell-edge performance by jointly processing cell-edge users' data as a unique entity. There are two classifications for joint processing; the first one is the Joint Transmission (JT), and the second one is the Dynamic Cell Selection (DCS). In Joint Transmission the same resource block of the PDSCH is transmitted from multiple cells associated with a UE-specific demodulation Reference Signal (UE-RS) among coordinated cells [12]. If the cells' joint transmission was time synchronized it is called coherent JT, if it is not synchronized it is called non-coherent JT. Also in Dynamic Cell Selection the PDSCH data has to be available at many cells. However, it is only sent from one cell at a given time in order to reduce interference [15].

In CoMP reception in the uplink, the Physical Uplink Shared Channel (PUSCH) is received at multiple cells where maximal ratio combining (MRC) is used. Two main schemes could be used, multi-point reception with Interference Rejection Combining (IRC) and the multi-point reception with coordinated scheduling. In the IRC scheme, the same resource block at the PUSCH is used by simultaneous transmission of multiple UEs, while the received weights are generated so that the received SINR or signal power is maximized at the central eNodeB in CoMP reception. In the multi-point reception with coordinated scheduling scheme, the received PUSCHs at multiple cell sites are combined by the use of Minimum Mean Squared Error (MMSE) or Zero Factoring algorithm. Both schemes improve the cell-edge user experience due to the increase of the signal power [12].

In 3GPP Release 8 the generation of the Demodulation Reference Signal (DMRS) embedded in two defined SC-FDMA symbols in an uplink sub-frame depends on the physical cell identity (PCI) which is derived from the downlink. A fundamental change to CoMP in the LTE uplink is the introduction of virtual cell ID's, which means that the macro cell and small cells are using the same cell identities in the case of heterogeneous network deployments [15].

There are further enhancements to the CoMP In 3GPP release 12/13 in both ideal and non-ideal back-haul scenarios. In ideal back-haul scenarios, the introduction of CSI-RS based RSRP measurement, uplink sounding and power control enhancement. While in the non-ideal back-haul scenarios, schemes will be developed to deal with the limitation of back-haul when using CoMP in order to get higher cell edge throughput and more efficient mobility management [7].

In [6], they propose a CoMP protocol stack and an architecture. They considered a logical CoMP cooperation entity that performs the resource joint scheduling procedure to allocate the same radio resources to CoMP cooperating cell to provide services to UE, and it is connected to each eNodeB by a C1 interface as the Figure 6 shows. They also proposed a signalling flow chart among this entity and eNodeBs as described in Figure 7. The CoMP serving gateway could serve the synchronization transmission in multiple eNodeBs as a central process node, and it will be connected to the eNodeB using a C2 interface. There must be a synchronization protocol to keep content synchronization and packet loss discovery as shown in Figure 8 or in Figure 9. The CoMP serving gateway could be an isolated node in the network or in eNodeB itself, in this case the use of S1 or X2 interface will be used as in Figure 10.

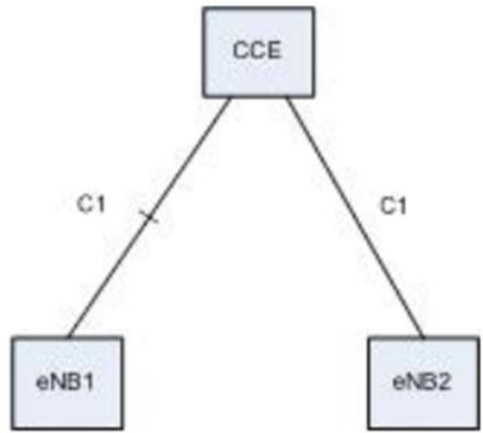

Figure 6. The CoMP cooperation entity [6]

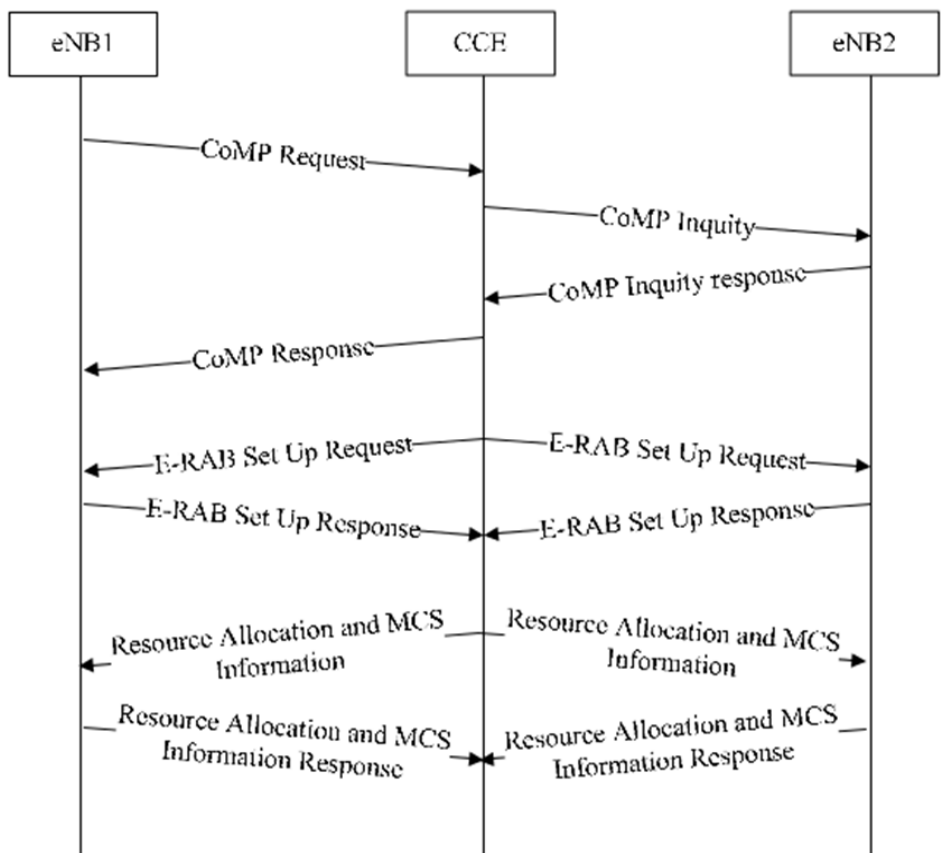

Figure 7. CoMP signalling flowchart proposed by [6]

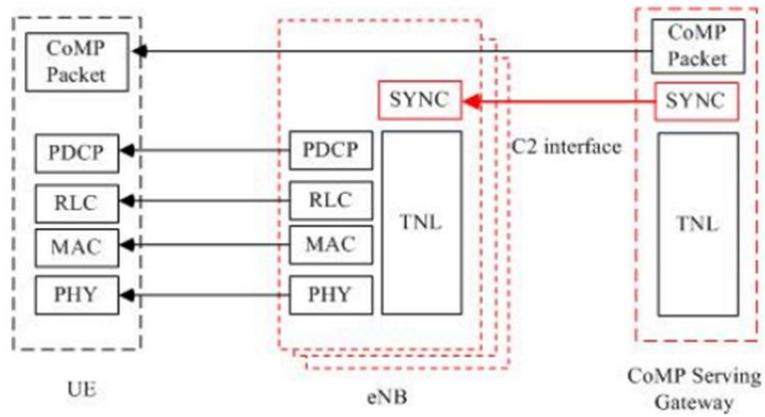

Figure 8. The AS process when it is located in the eNodeB [6]

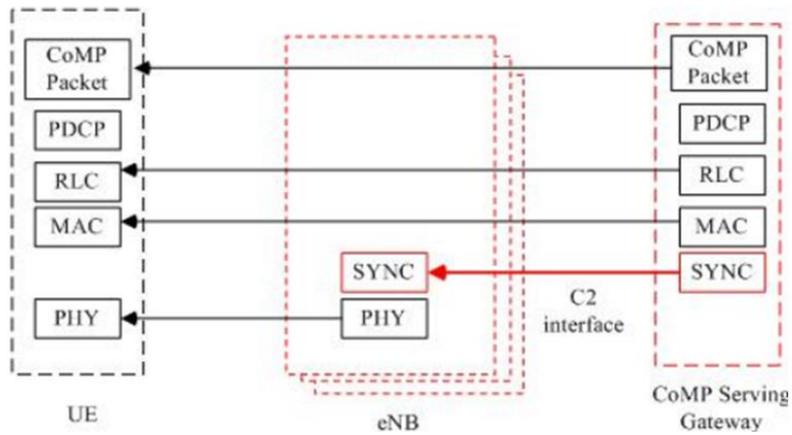

Figure 9. The AS process when it is located in a central node [6]

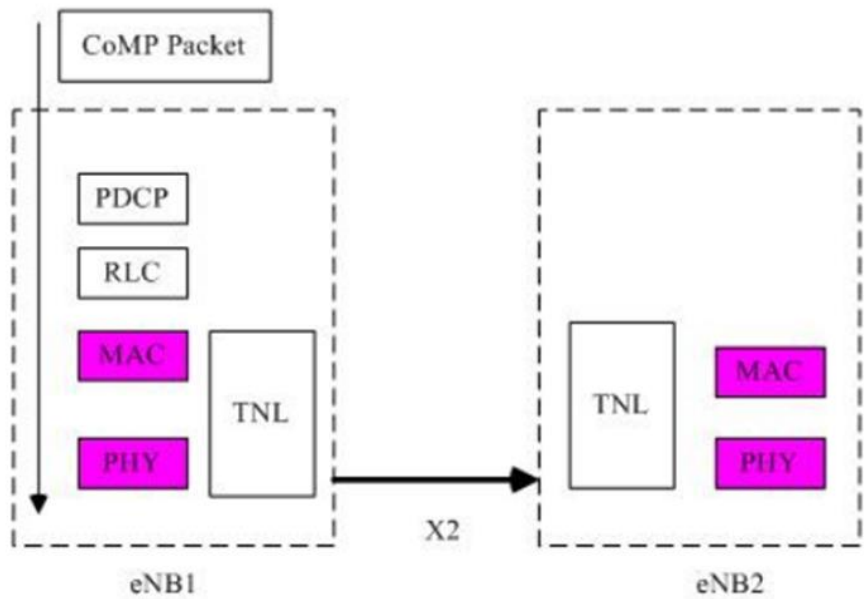

Figure 10. Data transmission by the X2 interface [6]

International Journal of Wireless & Mobile Networks (IJWMN), Vol.14, No.4, August 2022

## 5. RELAYING & BACK-HAULS

Relaying is a method of improving the network coverage in difficult conditions, such as to improve urban or indoor throughput, or to extend coverage in rural areas, or to add dead zone coverage [1]. This coverage improvement could be accomplished by deploying more base stations wired, connected to the rest of the network. However, relays are more attractive choice, since they have a shorter time deployment and no need to deploy a specific back-haul [3].

Two types of relays are being deployed in 3GPP Release 8, Amplify-and-forward relays, Decode-and-forward relays. In the Amplify-and-forward relays, the signal is simply amplified and forwarded, this type of relays are being used in coverage holes. In the Decode-and-forward relays the signal is decoded and re-encoded before re-transmission. This type of relays do not amplify the noise which makes it a suitable deployment choice in low-SNR environments [3].
In 3GPP Release 10 the concept of relaying has evolved to a level in which the relay nodes can connect to the donor cell's base station in two ways, the difference is whether to use the same channel or not. In the case where the donor cell's base station to relay node link "Back-haul link" operates on the same channel as a relay node to users' links "Access links" is called In-band relaying. If the back-haul link and access links are operating on different channels, the relaying is called an out-band relaying [1].

The interference in out-band relaying could be avoided in the frequency domain by sufficiently separating the back-haul link from the access links. Therefore, it could be operated on 3GPP Release 8 air interface without any enhancements. While in in-band relaying, unless the transmission on the back-haul link and the access link are being separated in the time domain, a proper antenna arrangements are needed to avoid the interference [3].

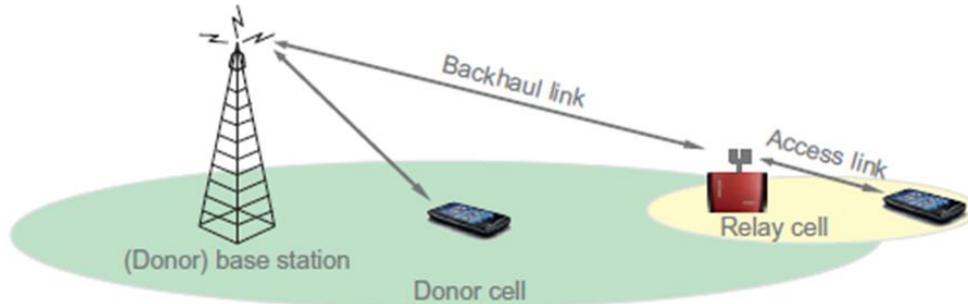

Figure 11. Access and Back-haul links [3]

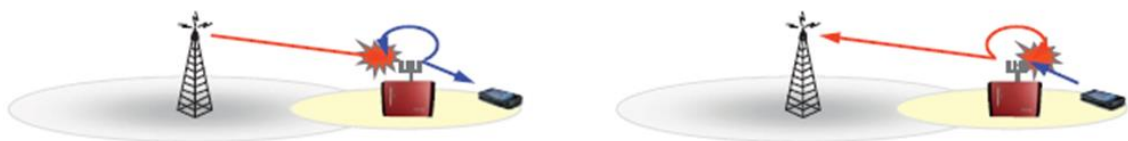

Figure 12. Interference between access and back-haul links [3]

The traditional way of scheduling radio resources to each user at the base station is based on the channel conditions and the service type for each user. This scheduling mechanism will become more complicated if the relay node was introduced, this is because the donor base station is only aware of the channel conditions of the back-haul link, yet it is not aware of the access link's channel conditions. A straight forward approach to solve this issue is by forwarding the access link's channel conditions to the donor base stations. However, this solution is very complicated, overhead feedback is very high, and it will contradict the LTE specifications because the relay node is no longer a same control entity as a base station [23].

In [23], they solved the above issue by proposing a two-step back-haul resource allocation scheme for an in-band relaying system based on relay buffer level. They used the relay buffer level and the amount of data waiting for downlink transmission across all logical channels as an indicator of the donor base station to decide the resource demand for each relay node taking into account the back-haul link quality and the access link quality. They evaluated their scheme which they call "Buffer-based" scheme with two other schemes, the "back-haul exclusive" scheme and the UE-based scheme. In the back-haul-based scheme, the relays and macro UE in a macro cell each get half of the time-frequency resources for transmission. In the "UE-based" scheme, the number of UEs in each relay is used as criteria in partitioning the resources in back-haul sub-frames. Their results showed an improvement in cell-average and cell-edge users.

## 6. RADIO RESOURCE MANAGEMENT FOR EICIC TECHNIQUE IN LTE-A

In [11], the authors aim at describing the paradigms for the design and operation of heterogeneous cellular networks that are composed of pixels, photocells and relays. Their particular focus was on cell splitting, range expansion, semi-static resource negotiation on third-party back-haul connections, and dynamic interference management for QoS via over the air signalling.

The authors of [11] introduced the differences between the deployment of macro-cells and the heterogeneous cells in terms of interference management and traffic load. Then they briefly listed and commented on the related work found in the literature. Also, the authors explained the concept of cell splitting and range expansion. They demonstrated the effects of Range Expansion to the user's throughput. They did this by considering an example in which a mobile may associate with a Pico base-station even though the received power from the closest macro base station on the downlink is higher. Their results showed that the user's throughput is increased by the range expansion. However, this could lead to a higher interference from a macro base station at the mobile which is associated with the Pico base station. They also demonstrated the users' association rate to Macro and Pico base stations by the use of bar graphs.

Also in [11], the authors stated that their focus of the interference management techniques was on the Downlink. They started by representing their system model with mathematical denotations. Then they classify the interference management techniques according to the timescale of coordination into two main classes. The semi-static Interference Management which they explained in part four and the Fast Dynamic Interference Management which they explained in last part of their paper. Also, the authors explained the Joint Association and Semi-static Resource Allocation. They started by defining their optimization problem in maximizing the sum of utilities of average rates for mobiles across cells, and they defined the formula of spectral efficiency for serving mobile $j$ on sub-band $m$. Then, they stated that the interference neighbourhoods could be represented by a graph $G(S)=(B, E(S))$, with node set $B$, and an edge set $E(S)$ where two base stations $i$ and $i_{prime}$ are connected by an edge if $i \in I(i_{prime}, S)$ or $i_{prime} \in I(i, S)$. Then, they adapted their optimization problem to a given association and a transmission power. Then, they reached to a numerical complexity result to their optimization problem of $O(N^M P^{RN})$ specifically, for N base-stations, M mobiles, R sub-bands, P power levels.

Also in [11], the authors thoroughly explained Dynamic Interference Management algorithms. These algorithms focus on networks of femtocells with a Closed Sub-carrier Group (CSG) in which each mobile could associate only with its owner's femtocell. It also prioritizes packets based on marginal utilities of average rates for best effort users and head-of-line delay for delay QoS users in each subframe. They started this part by explaining how the base-station could use an optimal modulation scheme based on the signalling messages exchanged with the User Equipment (UE) in a process that is called Rate Adaptation. Then they explained their prioritization formulas that they used to prioritize the main traffic types that they considered in their work, the Best Effort Elastic Flow, Delay QoS Flows, and a mix of them both. Then they explained Over-the-air coordination by stating the information that each base-station has to be aware of, and how interference coordination across cells occurs through the exchange of the *CoordReq* and the *CoordMsg* messages. Then they explained the two resource selection schemes, the independent and the coordinated sub-band selection. Then they explained how the base-station computes the transmission power to a mobile in its cell as a function of the *CoordMsgs* received from victim mobiles, in which this power is indicated on the downlink pilots to the victim mobiles. Then they explained how they computed the minimal interference in vector calculations in order to make it minimal. Then they did a numerical comparison between the optimal centralized scheme, their heuristic, and for reuse one scheme where a transmitter transmits on randomly selected sub-bands depending on the amount of traffic in the buffer, and they displayed their results in line graphs. As regards of the reuse one scheme, their results stated that around 70% of the delay sensitive traffic flows have 90th percentile packet delay of more than 200ms while about 30% full buffer users have rated less than 10 Kbps, which they explained it by the large amount of interference in a femtocell environment. As regards to both the centralized scheme and the distributed heuristic, the use of them leads to a substantial improvement in the performance of the full buffer and delay sensitive users. Their results stated that the performance of the distributed heuristic is not as well as the optimal algorithm. However, when they did one round of the very few bits of information exchange, it leads to 90th percentile delays of less than 25 ms for almost all QoS users, and rates above 5 Mbps to 90% of full buffer users.

The authors of [8] proposed a comprehensive ABSF framework that is able to mitigate the interference in HetNet environments that is comprised of macro-cell and Femto-cells. They started by introducing the definition of Enhanced Inter-Cell Interference Coordination (eICIC) and the categories of the solutions found in literature that are being used to mitigate interference between macro-cells and Femto-cells in HetNet environment. Then they overviewed the choice ABSF mechanism. Then they summarised the contribution that their proposed ABSF framework provides, such as providing a complete tracking procedure for the MUEs to mark and unmark their victim state, and other contributions which are listed in the paper. Then they explained their system model by explaining the network architecture, the communication nodes, and the assumptions of their scenario and how they considered the DL of an LTE-A HetNet environment.

Also in [8], the authors thoroughly explained their proposed ABSF framework and they stated that it was developed in order to assess the potential of ABSF technique in resolving interference problems in HetNet environments. They started by explaining how they track MUEs and how they mark the ones with an SINR that is below a certain level as Victim MUEs (VMUEs). Then they explained how the process of activating and deactivating of ABSF mode in an Aggressor HeNBs "which is the HeNB that is located near an MUE and generates strong interference in the downlink" is done. Then they explained the process of selecting the ABSF pattern, and this is important because different ABSF patterns will allow the VMUEs differential interference-free bandwidths. Then they explained the coordination that occurs between MeNB, HeNB and MUEs through the exchange of signalling messages "CQI Report Control Message, RSRP Measurement Report Control Message, and the ABSF Mode Triggering Control Message". Then they explained the enhancements that they did to the downlink scheduler at the Macro eNB by the use of a scaling metric that they defined in a formula. Then they explained how they tracked the SINR during the ABSF using the Kalman filter.

Also in [8], the authors explained how they evaluated the performance of their system by the use of LTE-Sim and they specified the parameters of their scenario. They studied different issues. For both the ABSF and the normal mode, they studied the aggregate throughput of the macro-cell and the aggregate throughput of the VMUEs under different input load. Their results showed that the ABSF mode had a higher aggregate throughput of the macro-cell, and it had an outstanding higher aggregate throughput of the VMUEs. They also studied the aggregate throughput of the Femto-cells under different input load for both the ABSF and the normal mode. Also, their results showed that both modes had a constant aggregate throughput, but the ABSF had a higher value. They also studied the SINR for a VMUE during ABSF and normal subframes. They also studied the fast response and accurate tracking of the SINR using the Kalman filter for both the ABSF and the normal mode, their results showed a similar performance. Finally, they drew conclusions and forecasts future works which will focus on developing a scheme for the ABSF pattern selection and to propose an integrated ABSF-PC (Power Control) scheme and evaluate its performance.

## 7. RADIO RESOURCE MANAGEMENT FOR CoMP TECHNIQUE IN LTE-A

In-order for the scheduling process to take place in LTE-A in the presence of CoMP, the cellular cells' sectors have to be clustered. There are three clustering approaches as shown in Figure 13; the fixed clustering approach as shown at the left of the figure, the pure UE-specific clustering approach as shown in the middle of the figure, and the semi-UE-Specific clustering approach as shown at the right of the figure [24].

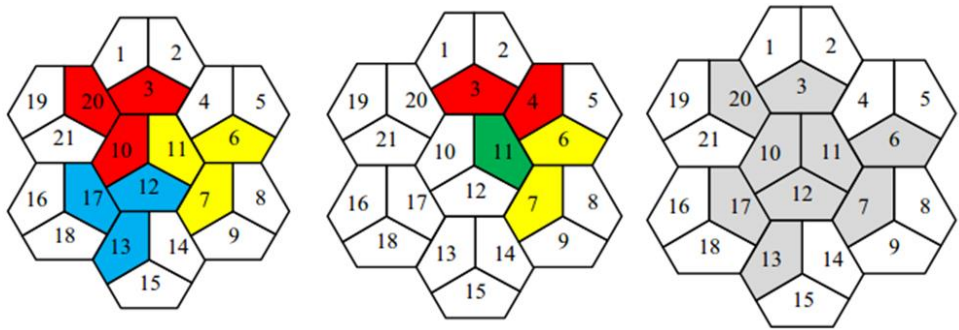

Figure 13. The three clustering approaches; the fixed clustering approach as shown at the left of the figure, the pure UE-specific clustering approach as shown in middle of the figure, and the semi-UE-Specific clustering approach as shown at the right of the figure [24]

In the fixed clustering approach as shown at the left of Figure 13, the network is divided into non-intersecting coordinated clusters, in which the UE in a sector, will only be served by the sectors in the same cluster. For example, if the UE existed at the edge of sector 11, it would only be served by the coordinated sectors 11, 6, and 7 despite the fact that it would get a better service if it was served by the coordinated sectors 10, 11, and 12. As this example illustrated that the use of this approach has limitations, and doesn't always provide the best service.

In the pure UE-specific clustering approach as shown in the middle of Figure 13, the choice of selecting which cluster of coordinated sectors to be served by following the UE preference that is based on the Reference Signal Received Power (RSRP). For example, let us consider a scenario where two cell-edge UEs being separately served by the two intersecting coordinated clusters (3,4,11) and (6,7,11). Clearly, they would provide the highest throughput gain, but the complexity of implementing this approach is very high since it requires scheduling radio resources across all eNodeBs.

In the semi-UE-Specific clustering approach as shown at the right of Figure 13, there is a similarity with the pure UE-Specific, but the cluster is a sub-set of a larger cluster instead of the whole network. For example, if a the cell-edge UE exists in sector 11, it can choose any two other sectors within the grey area. This approach can reach a balance between gaining throughput and implementation complexity, and this is because it only requires scheduling radio resources across the eNodeBs in a larger fixed cluster instead of scheduling them across all the eNodeBs.

According to [24], in the case of fixed clustering, the use of the RR scheme of the CoMP system is almost similar to it in the non-CoMP system, except that there is a different treatment while allocating the RBs to cell-edge UEs and cell-center UEs. The frequency band is divided into two parts, one to perform the CoMP operation for cell-edge UEs, and the other to perform the non-CoMP for cell-center UEs. After dividing the frequency band, the RBs of CoMP frequency zone are assigned to all clusters' cell-edge UEs equally and in a circular order, and meanwhile assigning RBs of the single cell frequency band to all sector's cell-center UEs equal and in circular order. This RR scheduling approach can't be applied directly to the UE-specific and the semi-UE-specific cases due two main reasons, the conflict resources and the residual resources. The conflict in resources happens when allocating a RB that belongs to an overlapped sectors to two different UEs that belongs to two intersecting coordinated clusters. The residual resources happens when there are some RBs left after allocating, which is also caused by the overlap of coordinated clusters.

In [24], they solved those two conflicts by modifying the RR scheduling approach as shown in Figure 14. They tested their scheduling approach by comparing it to two other schemes in terms of cell spectral efficiency (bit/s/Hz) and cell-edge UE spectral efficiency (bit/s/Hz), one is a non-CoMP system with traditional RR scheduling, the other is a fixed clustering CoMP system with traditional RR scheduling. In terms of cell spectral efficiency, their proposed RR scheme based on pure UE-specific CoMP achieved a 6.97%, increasing gain when compared to the non-CoMP approach, and the RR scheme based on fixed clustering CoMP achieved a 3.03% increasing gain when compared to the non-CoMP approach. In terms of cell-edge UE spectral efficiency, their proposed scheme achieved a 22.46%, increasing gain when compared to the non-CoMP approach, and the RR scheme based on fixed clustering achieved a 6.52% increasing gain when compared to the non-CoMP approach.

The scheduling procedure of uplink systems In CoMP joint scheduling is done as shown in Figure 15. The base stations handle the joint scheduling and resource arrangement based on the calculated CSI that is obtained by the uplink sounding reference signal, then this information is exchanged between the coordinated points. Based on these joint scheduling results that are carried on the downlink control channel, UE is scheduled and the coordinated base station's joint will receive the data from the scheduled UE, and also the sounding reference signal is transmitted. The joint scheduling scheme is suboptimal for the case of MU-MIMO system. This is due to the large number of UEs. However, the use of it will decrease the processing complexity [10].

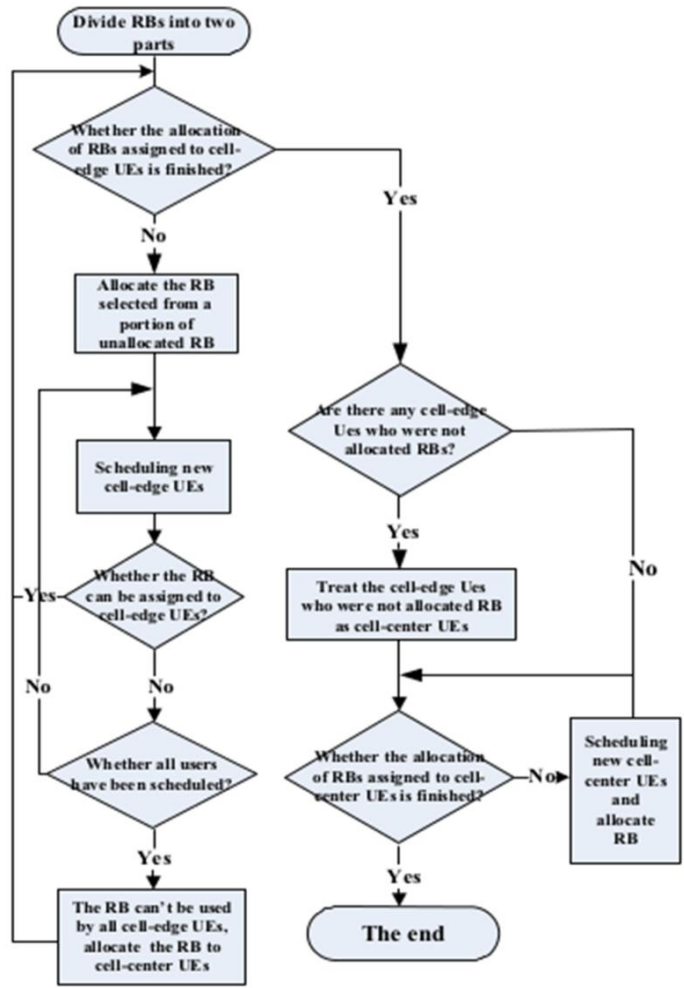

Figure 14. The modified Round Robin scheduling approach proposed by [24]

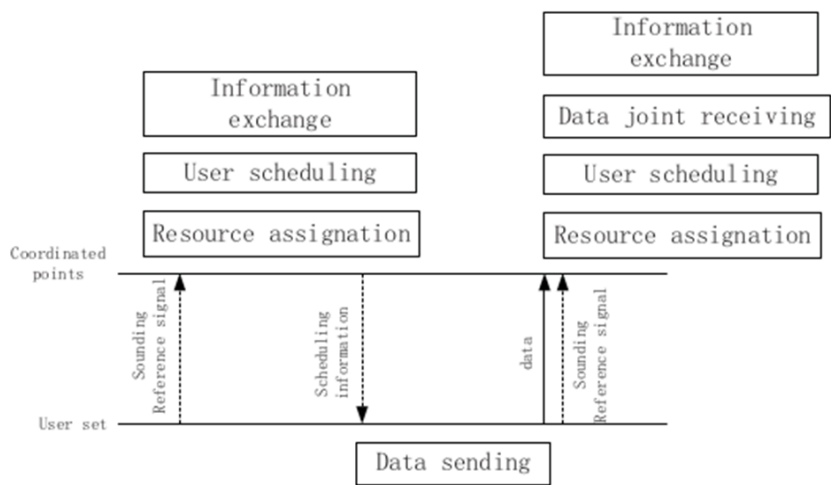

Figure 15. Scheduling procedure of uplink systems [10]

In [10], they have proposed a joint scheduling scheme for MU-CoMP. Their scheme is designed in two parts that combines the PF criteria and the orthogonal criteria. The first part, the PF scheduling algorithm chooses the primary user within the serving cell, and denotes it as the target UE. The scheduler chooses the target UE after comparing the value of estimated instantaneous data transmission rate over the average transmission rate for each user in the cell, and then it chooses the one with the higher value. After that the coordinated base stations are selected based on the reference signal that is received by the target UE. Then, within the scope of coordinated base stations all the UEs is collected to form a set, this set includes a sub-set that contains the UEs within the maximum PF values, and to be prepared to pair with the target UE. In the second part, the orthogonal factor for each user in the sub-set is calculated, and then chooses the one with the highest value to be paired with the target UE. The orthogonality is useful in reducing the interference among the paired UEs.

## 8. RADIO RESOURCE MANAGEMENT FOR RELAYING AND BACK-HAULS TECHNIQUE IN LTE-A

The Addition of relay nodes in LTE-A networks will raise new challenges on the resource allocation concerning how to split them at Donner eNodeB (DeNB) between the macro-access link and the backhaul link. According to [4], Two algorithms could be used. One algorithm is called the Fair Resource Unit (Fair-RU), and the other one is called the Fair-Throughput (Fair-TP) strategy.

The Fair Resource Unit (Fair-RU) is a simple and straight forward approach which aims at partitioning the resources according to the following formula:

$$\epsilon = \frac{N_R}{N_R + N_M} \cdot \delta$$

where $N_R$ is the number of relay-attached UEs (R-UEs), $N_M$ is the number of macro eNB-attached UEs (M-UEs) which are always served directly by the DeNB, where $\delta$ is a factor that scales $\varepsilon$ depending on the total number of available resources, and its value differs according to either the in-band relaying, or the out-band relaying. In the case of the in-band relaying, $\delta = N_f / p$, where p is a fixed number of MBSFN frames in a period of $N_f$ consecutive sub-frames. In the case of out-band relaying, $\delta$ is defined as the ratio between the number of carriers that can transport the back-haul link and the total number of carriers used at the DeNB.

The Fair-Throughput (Fair-TP) strategy allows for dynamic partitioning. In the case of fixed number of MBSFN sub-frames, the algorithm calculates the aggregate average throughput on the back-haul link $\Phi_B$ and the aggregate average throughput $\Phi_M$ on the access link to all macro UEs. Within the non-MBSFN frames it is expected that the aggregate average throughput $\Phi_M$ will increase, while the macro users are scheduled, and that $\Phi_B$ will decrease. The aim is to keep allocating resources to the back-haul link until a ratio ($N_R/\Phi_B$) < ($N_M/ \Phi_M$), if it exceeded this value, radio resources will be allocated to macro UEs.

The use of Fair-TP is reasonable in the case of in-band relaying due the few transmission opportunities available for the back-haul link. However, it is not reasonable to use it in the case of out-band relaying due to carrier separation. Hence, the back-haul link could be included in the resource allocation of the whole macro cell using the regular PF scheduling algorithm in a process called the Ext-Prop-Fair.

In [4], they built a QoS scheduler that consists of a TD scheduler and an FD scheduler. Their network architecture that they run their simulation based on is shown in Figure 16, where RN stands for Relay Node, and NCE stands for NOMOR Channel Emulator. Their TD scheduler creates a candidate list, in which it will be forwarded to FD scheduler. The FD scheduler is responsible for allocating radio resources to users in the candidate list. The FD scheduler works by visiting radio resources one by one in order to allocate each of them to the user with the best metric, and it also calculates the average throughput of the allocated user on each resource unit and updates this value. This resource allocation process is repeated until there are no resource units or the transmitting buffer is empty.

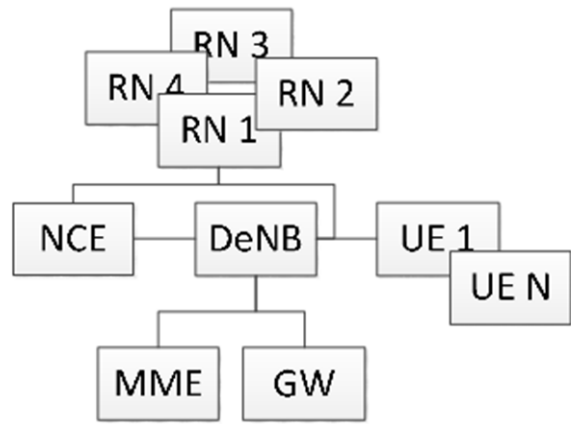

Figure 16. LTE-A test-bed architecture that was proposed by [4]

The deployment scenario that was used in [4] consists of a single macro cell that contains one DeNBs and two RNs attached to it and 25 UEs. UEs were separated into two groups, 21 M-UEs, and 4 rules in which each RN serves 2 R-UEs. Two 5 MHz carriers were used in the case of out-band relaying, and one single 10 MHz carrier in the case of in-band relaying. The traffic generator that they used is IP-based, and they implemented it in the PDCP layer of the DeNB protocol stack. They defined two types of traffic with different QoS requirements. One was defined as VoIP traffic that was emulated by a Constant Bit Rate (CBR) traffic of 128 kbps with a maximum end-to-end delay of 100 ms and a service priority of 2. The Other one was defined as Video Streaming traffic that was emulated by a CBR traffic of 256 kbps with a maximum end-to-end delay of 300 ms and a service priority of 5.

In [4], when they compared the resource partitioning results for out-band relaying to DeNB only with carrier aggregation, they found that a possible gain could be achieved when using any of the three schemes, the Fair-RU, Fair-TP, and the Ext-Prop-Fair. However, Ext-Prop-Fair seemed to be the better solution because of its unified framework and best performance in the low and medium range. Also, when they compared the resource partitioning results for in-band relaying to DeNB-only without carrier aggregation a similar result of possible gain could be achieved when using any of the three resource partitioning schemes. However, they found out that the Fair-RU penalizes the M-UEs because of over-compensation of back-haul link, so the Fair-TP seemed to be a better choice.

In [4], they compared their QoS aware scheduler to the conventional static resource partitioning in terms of throughput and delay for M-UEs and R-UEs, they tested both VoIP and Video streaming traffics. In terms of throughput, satisfied M-UEs "users who received at their minimum pre-configured bit rate for their traffic type" was increased from 8.1% to 88.8% for video streaming traffic type, and it also increased from 80.0% to 99.4% of VoIP traffic. In terms of delay, satisfied M-UEs "users who received their packets within their defined delay" increased from 74.7% to 95.3% of video streaming traffic, and it also increased from 90.5% to 99% of VoIP traffic. In the case of R-UEs, both schedulers had a 100% satisfied user of the two traffic types.

The authors of [13] proposed a resource allocation strategy that was able to effectively support and satisfy a QoS level for real-time applications LTE-A network with the existence of relay nodes, this was accomplished by extending their previous proposed Two-Level Scheduler approach in [14]. They adapted their Two-Level Scheduler to the relay technology by applying it to both the DeNB and the relay, and also by splitting the maximum delay on the access network from the DeNB to the terminal $\tau_i$ into two parts as follows:

$$\tau_i = \tau_{i,DeNB\text{->}relay} + \tau_{i,relay\text{->}UE}$$

Where $\tau_{i,DeNB\text{->}relay}$ and $\tau_{i,relay\text{->}UE}$ are the delay on the back-haul and on the user links, respectively, as shown in Figure 17.

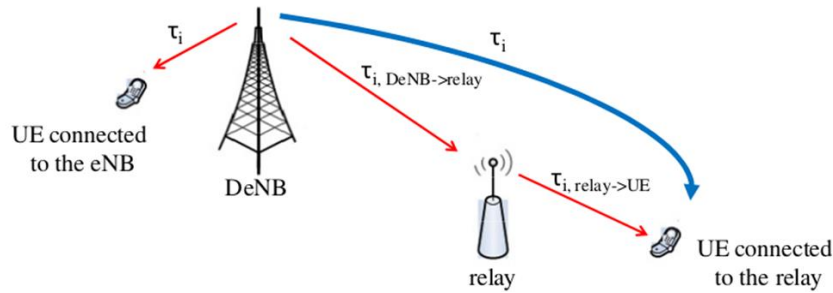

Figure 17. Target delays in a multi-hop connection [13]

In [13], they used the LTE-Sim framework to test the performance of their strategy in several network scenarios that involve a variable number of users hosting real-time multimedia applications and best effort flows. And also to compare their strategy with respect to Proportional Fair and Logarithmic strategies. The performance index that they based their comparison based on depended on the type of the traffic. In the case of multimedia applications such as Voice over IP (VoIP) and video, they measured the Packet Loss Ratio (PLR). And for the case of best effort traffic, they measured the goodput "which defined as the rate of useful bits they successfully transmitted during the whole simulation" and the fairness.

The simulation results of [13] for the multimedia applications showed that the PLR increased proportionally with the number of users, due to increased network load. And for the best effort, they noted that the goodput decreased as the number of users increased. Their results showed that their approach was able to reduce PLR for multimedia applications when compared to the PF and LOG approaches, but with the expense of a slight degradation in the throughput of the best effort flows. And their results also showed that both LOG rule and PF approaches obtained a higher goodput, due to the fact that they both provide a worse service to multimedia flows when compared to their approach, thus leaving a higher share of bandwidth for best effort traffic. In order for them to gain more accurate comparisons, they evaluated the Jain Fairness Index, finding that it was higher than 0.8 in all considered conditions. Which led to a conclusion that all approaches were able to provide high levels of fairness to best-effort applications.

## 9. CONCLUSION

This survey paper has provided a detailed explanation of the concepts of some of the techniques that are used with LTE-A. These techniques are the Inter-cell Interference Coordination (ICIC), enhanced ICIC (eICIC), Coordinated Multi-point operation (CoMP), Relaying and Back-hauls. Then, it summarised the proposed Radio Resource Management (RRM) approaches that were proposed in the literature for these techniques within an LTE-A network deployment. The importance of this paper, is laying the foundation of understanding the technologies which will be further studied and modelled as a part of the 5G network and beyond 5G.

### CONFLICTS OF INTEREST

The authors declare no conflict of interest.

## APPENDIX A: LTE CHANNELS

LTE Channels are classified into three main types; logical channels, transport channels, and logical channels. LTE Logical channels are shown in Figure A1. LTE Transport channels are shown in Figure A2. LTE physical channels in downlink and uplink are:

*Physical Downlink Shared Channel (PDSCH)*, it is responsible for carrying users' data, and it uses different modulation types, such as QPSK, 16-QAM, and 64-QAM [2].

*Physical Broadcast channel (PBCH)*, it is used for sending cell-specific system identification and access control every 40 ms, and it uses QPSK modulation [2].

*Physical Control Format Indicator Channel (PCFICH)*, it is used for indicating the number of OFDM symbols used for the transmission of control channel (PDCCH) information in a sub-frame by a parameter that has the values from 1 to 3, and it uses QPSK modulation [2].

*Physical Downlink Control Channel (PDCCH)*, it is used for informing users about resource allocations of PCH and DL-SCH, and Hybrid ARQ information related to DL-SCH. It Carries the uplink scheduling grant, and it uses QPSK modulation [5].

*Physical Multi Cast Channel (PMCH)*, it is used to carry multi-cast information which are sent to multiple users simultaneously, it uses different modulation types such as QPSK, 16-QAM, and 64-QAM [2].

*Physical Hybrid ARQ Indicator Channel (PHICH)*, it is used to carry the ACK/NAKs in response to uplink transmission that confirms the delivery of data or request the retransmission of data blocks received incorrectly [2].

*Physical Uplink Control Channel (PUCCH)*, it is used to carry uplink control information such as channel quality indication (CQI), ACK/NAK responses of the users to the HARQ mechanism, and uplink scheduling requests. It is never transmitted with PUSCH data [2].
*Physical Uplink Shared Channel (PUSCH)*, it used to carry the UL-SCH, and it uses QPSK, 16-QAM, and 64-QAM Modulation [5].

*Physical Random Access Channel (PRACH)*, it is used to carry the random access preamble which is generated from Zadoff-Chu sequences with zero correlation zone, generated from one or several root Zadoff-Chu sequences [5].

International Journal of Wireless & Mobile Networks (IJWMN), Vol.14, No.4, August 2022

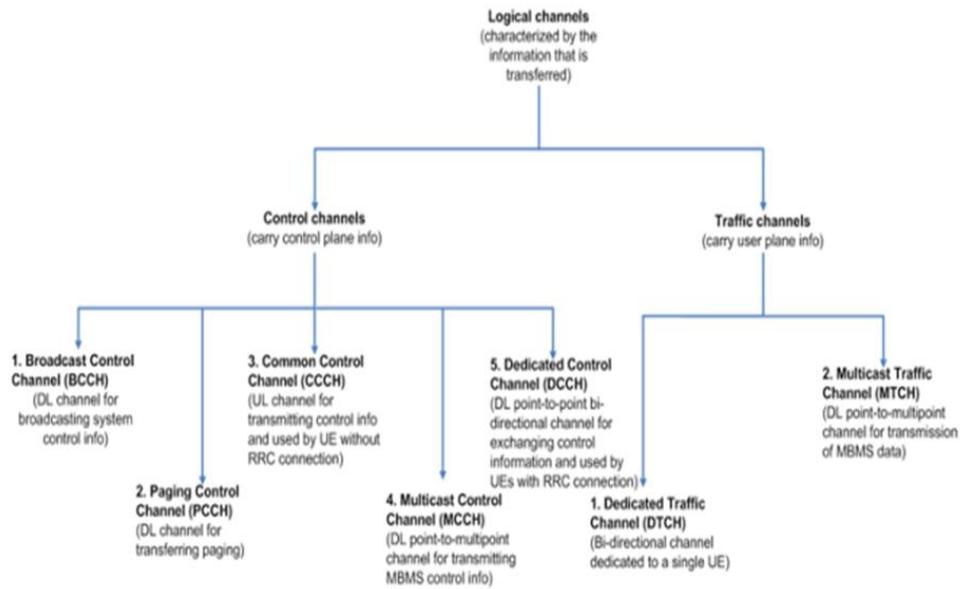

Figure A1. LTE logical channels [9]

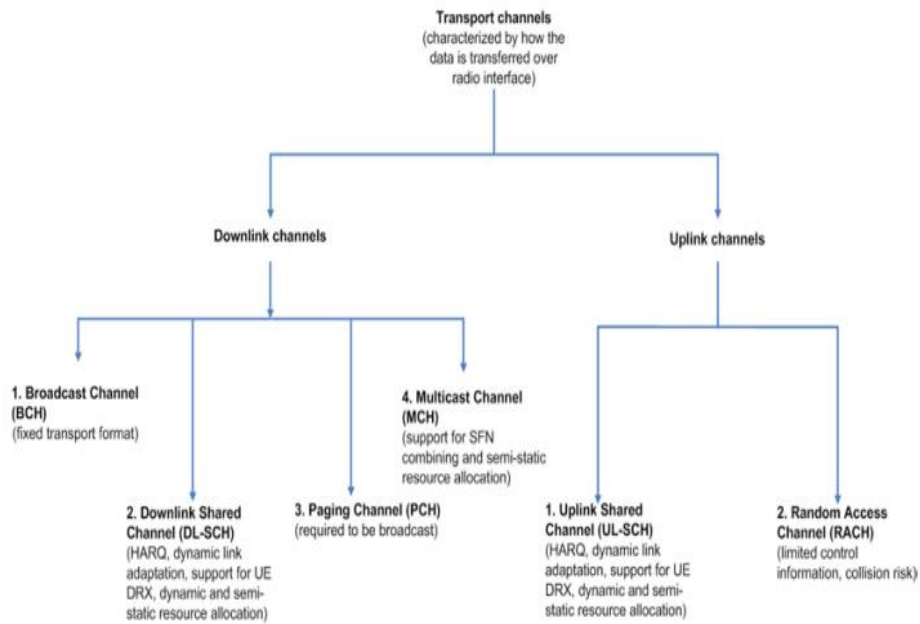

Figure A2. LTE transport channels [9]

The mapping between all these channels in downlink and uplink are shown in Figures A3, A4 respectively.

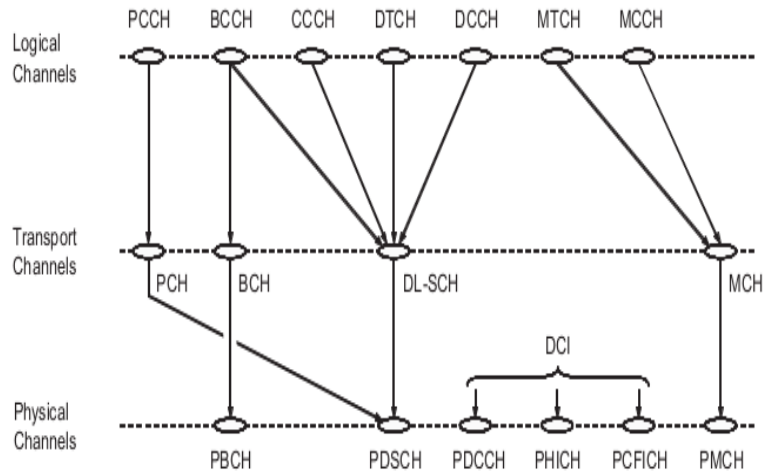

Figure A3. Downlink channels mapping [3]

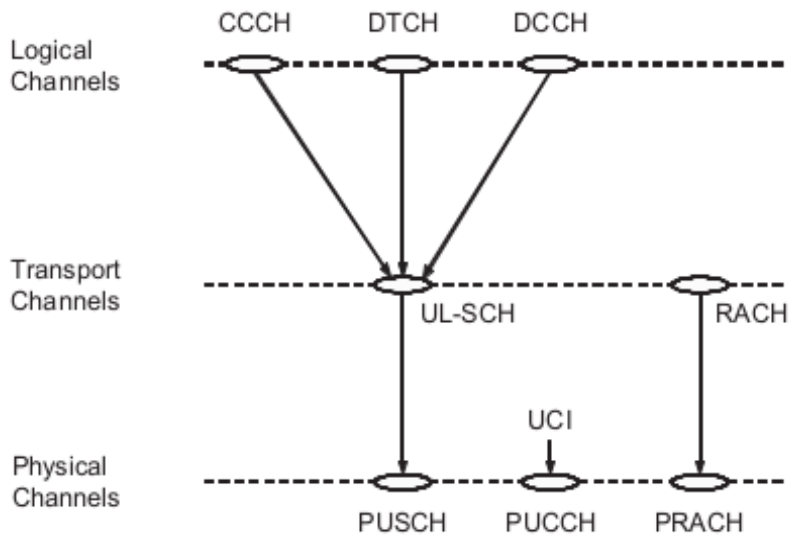

Figure A4. Uplink channels mapping [3]